\documentclass[pra,twocolumn,amssymb,aps,floatfix,nobibnotes]{revtex4-2}
\usepackage{graphicx}
\usepackage{physics}
\usepackage{amsmath}
\usepackage{mathtools}
\usepackage{dcolumn}% Align table columns on decimal point
\usepackage{bm}% bold math
\usepackage{hyperref}% add hypertext capabilities
\usepackage{braket}
\usepackage{color}
\usepackage{xcolor}
\usepackage{enumerate}
\usepackage[normalem]{ulem}
\usepackage{tikz}
\usepackage{dutchcal}
\usepackage{cancel}
\usepackage{tcolorbox}
\usetikzlibrary{arrows}
\usetikzlibrary{calc}
\newcommand{\tikzmark}[1]{\tikz[overlay,remember picture] \node (#1) {};}
\tikzset{square arrow/.style={to path={-- ++(0,-.25) -| (\tikztotarget)}}}
\newcommand\myline[1][]{%
	\,\tikz[baseline]\draw[very thick,#1](0,-\dp\strutbox)--(0,\ht\strutbox);\,}
\newcommand{\be}{\begin{equation}}
\newcommand{\ee}{\end{equation}}
\newcommand{\bea}{\begin{eqnarray}}
\newcommand{\eea}{\end{eqnarray}}

\begin{document}

\title{Structure of the nearly-degenerate manifold of lattice quasiholes on a torus}

\author{Z. Zeybek}
\affiliation{%
Universit\"at Hamburg,
Institut f\"ur Laserphysik, Luruper Chaussee 149, 22761 Hamburg, Germany}%
\author{R. O. Umucal\i lar}
\affiliation{%
Department of Physics, Mimar Sinan Fine Arts University, 34380 Sisli, Istanbul, Turkey
}%

\date{\today}
\begin{abstract}
We study the nearly-degenerate quasihole manifold of the bosonic Hofstadter-Hubbard model on a torus, known to host the lattice analog of the Laughlin state at filling fraction $\nu = 1/2$. Away from $\nu = 1/2$ and in the presence of both localized and delocalized quasiholes, the ratio between the numerically calculated many-body Chern number for certain groups of states and the number of states in the relevant group turns out to be constant for this manifold, which is also manifested in the density profile as the depleted charge of localized quasiholes. Inspired by a zero-mode counting formula derivable from a generalized Pauli principle, we employ a combinatorial scheme to account for the splittings in the manifold, allowing us to interpret some groups of states as the quasihole excitations corresponding to filling fractions lower than $\nu = 1/2$. In this scheme, the many-body Chern number of subgroups appears as a simple combinatorial factor.
\end{abstract}

\maketitle

\section{Introduction}

The interest in topological phases of matter has tremendously increased since the discovery of the quantum Hall effect \cite{vKlitz,Tsui} and topological insulators \cite{FCIrev}. The rich phenomena revealed in this context, such as conductance quantization \cite{TKNN}, chiral edge states \cite{Halperin}, and excitations with fractional charge \cite{Laughlin} and exchange statistics \cite{FracStat,fqhStat, Haldane1991}, are theorized to be instrumental for a wide range of applications such as topological quantum computing \cite{QComp} and novel material design \cite{FCImat,PhotonicMat}. 

Recent developments in quantum simulation with ultracold atomic \cite{coldExp} and photonic systems \cite{photonExp} have created an alternative environment for investigating topologically ordered states such as fractional quantum Hall (FQH) \cite{BosonicFQH1,BosonicFQH2,Sorensen,HafeziTorus,photonicFQH} and fractional Chern insulator (FCI) states, the latter being exclusively studied for lattices \cite{RegnaultFCI,FCIexample,BernevigRegnault2012,NonAbelianFCI,FCIreview,Nielsen2015}. Validating the topological nature of states harbored in a given system is an essential component of such investigations, in which one typically looks for specific signatures such as charge fractionalization \cite{FractionalCharge}, braiding statistics \cite{QHLiu,OnurTTN} and matching of the excitation and/or the entanglement spectra of ground states \cite{RegnaultFCI,FCIexample,BernevigRegnault2012,NonAbelianFCI} with known results. Counting quasihole (zero-mode) states in an FQH-type system on a torus is considered to be a viable method highlighting the topological nature of the problem. Counting formulas for certain Abelian and non-Abelian quasihole states have already been provided either using a generalized Pauli principle \cite{RegnaultFCI,JACK} or in the thin-torus limit \cite{ThinTorusCount}. 

In this paper, focusing on the Hofstadter-Hubbard model of repulsively interacting bosonic particles in a finite lattice with torus boundary conditions, which is known to host the discrete version of the $\nu = 1/2$ Laughlin state \cite{Sorensen,HafeziTorus}, we scrutinize its nearly degenerate quasihole manifold with the guidance of a zero-mode counting formula valid for a generic Laughlin filling fraction $\nu = 1/r$. Based on numerical calculations, we find that, different from the continuum limit where the degeneracy is expected to be exact, there exist certain subgroups of states within a nearly-degenerate manifold which could be interpreted as the quasihole states existing around filling fractions lower than $\nu = 1/2$. We observe that the ratio of the numerically calculated many-body Chern number for the states in a subgroup to the number of states in that group (which we will call {\it Chern number per state} for brevity) is the same for all subgroups in the whole  nearly-degenerate manifold. This ratio also turns out to be equal to the ratio of the number of particles to the number of unpinned magnetic flux quanta in the system. Investigation of the density profile reveals that this ratio also represents the depleted charge at the position of pinned quasiholes. To classify the subgroups, we propose to use their many-body Chern numbers which can be expressed in a combinatorial form following a simple distribution rule. We believe that such a distribution rule and a compact form of the degeneracy formula we present, with its direct connection to fundamental topological concepts like the many-body Chern number and the fractional charge of quasiholes, could be seen as stimulating clues for new investigations towards a better understanding of the underlying quantum states, which may even include more exotic non-Abelian ones that can occur in a setting similar to ours \cite{Kapit Non-Abelian}.

\section{Generalized counting formula and the many-body Chern number} 

As quasihole excitations are among the unique defining features of FQH states, the matching of the number of low-lying states with that of Laughlin-type quasihole excitations for given system parameters can be considered to be a strong signal that the observed states resemble FQH states rather than competing ones like charge density waves \cite{RegnaultFCI}. 

The number of states lying below the many-body gap of an FQH-like system can be obtained both through a counting procedure in the thin-torus limit \cite{ThinTorusCount} and by using a generalized Pauli principle as detailed in Appendix A. In this latter approach, the total number of quasihole excitations of a $\nu_0 = N/N^{0}_\phi = 1/r$ state on the torus can be determined by applying the so-called $(1,r)$ generalized Pauli principle \cite{JACK}, in which one counts the number of distinct ways of distributing $N$ particles into $N^{0}_\phi$ orbitals obeying the condition that no more than a single particle is admitted in $r$ consecutive orbitals, as well as conforming with the periodic boundary conditions imposed by the torus. Using this method, we derived a formula to account for the ground-state degeneracy of a given system at an arbitrary filling fraction $\nu = N/N_\phi$ on the torus, assuming that the ground-state manifold corresponds to quasihole excitations of a reference Laughlin state with filling $\nu_0 = 1/r$:
\begin{align}
	&\mathcal{D}(N,N_d,N_\phi)=\frac{(N_d + N -1)!}{N_{d}!(N-1)!}\frac{N_\phi}{N}, \label{DEG}
\end{align}
where $N$, $N_\phi$, and $N_d = N_\phi-N/\nu_0$ are the number of particles, the number of flux quanta and the number of delocalized (unpinned) quasiholes \cite{QHLiu} in the system respectively. We note that although the above formula may not be considered new in the sense that it appears in similar forms in several other works \cite{ThinTorusCount,RegnaultFCI,FCIexample}, it is still original to the best of our knowledge, deriving its originality mainly from its simple present form with a direct connection to the many-body Chern number as we now explain. Our formula (\ref{DEG}) can be compactly written as $\mathcal{D} = \mathcal{C}/\nu$ with $\mathcal{C} = \binom{ N_d + N-1}{N_d}$ being the binomial coefficient. Most interestingly, in the cases we investigated for a lattice system via exact diagonalization, this combinatorial part $\mathcal{C}$ reproduces the numerically calculated many-body Chern number $C$ of the ground-state manifold so that the Chern number per $\mathcal{D}$-fold degenerate states is simply $\nu$. Similar to the case of a system with $\nu_0 = 1/r$ on the torus, for which the Chern number of the $r$-fold degenerate ground-state manifold equals one \cite{Wen,LLL}, this result suggests that $\nu$ might correspond to the fractional prefactor of the quantized Hall conductance in this case as well.

We note that for a filling fraction $\nu = N/N_\phi = k/r$, $k$ and $r$ being coprime, general translational symmetry arguments in continuum lead to an $r$-fold center-of-mass degeneracy on the torus \cite{Haldane} and the remaining degeneracy can be ascribed to relative translations \cite{BernevigRegnault2012}. For $k=1$, this latter degeneracy corresponds to the combinatorial part $\mathcal{C}$ of our formula, hinting at a potential relation between the many-body Chern number with the total number of states per momentum sector of the relative translation operators.

\section{Bosonic Hofstadter-Hubbard model} 

In this work we study bosons on a two-dimensional square lattice under a uniform perpendicular effective magnetic field, via considering the so-called Hofstadter-Hubbard (HH) model with the Hamiltonian:
\bea 
H = - t\sum_{\langle ij\rangle}\left(e^{i2\pi \phi_{ij}} 
c^\dag_{i}c_{j}+\text{h.c.}\right)+\frac{U}{2}\sum_i n_i(n_i-1),
\label{eq:Hofstadter Hamiltonian}
\eea
where $c^{\dagger}_i$ ($c_j$) is the bosonic creation (annihilation) operator at site $i$ ($j$) and  $n_i = c^{\dagger}_i c_i$ is the number operator. The strength of repulsive on-site interactions is given by $U>0$ and $t > 0$ is the hopping amplitude between nearest-neighbor sites. The magnetic field is characterized by the magnetic flux quantum per unit cell of the lattice defined as $\phi = \sum_{\square} \phi_{ij}/(2\pi)$, where the sum is carried out around a unit cell. For the non-interacting case with $\phi = p/q$, $p$ and $q$ being coprime, the energy band is split into $q$ sub-bands exhibiting the fractal Hofstadter butterfly \cite{Butterfly1,Butterfly2}. In the literature, there are quite a number of works (see e.g. \cite{Sorensen,HafeziTorus}) which have shown that the ground state of the interacting Hamiltonian (\ref{eq:Hofstadter Hamiltonian}) has a very large overlap with the Laughlin state for the filling fraction $\nu = N/N_\phi = 1/2$ under appropriate boundary conditions and in the so-called continuum limit $\phi \ll 1$. The robustness of this state has also been demonstrated through finite-size scaling studies which were performed both in the continuum \cite{Roy2016} and thermodynamic continuum \cite{Moller2018} limits, yielding an unambiguous many-body gap that scales as $1/q$.
 
Although we perform exact diagonalization for small hardcore bosonic systems, we employ the {\it lowest Landau level {\rm(LLL)} projection} method \cite{LLL} in real space to deal with larger systems. Working with larger systems helps us avoid boundary effects as much as possible, at the same time allowing us to maintain a certain number of flux quanta through the whole lattice while we reduce the flux quanta per unit cell to reach the continuum limit. An additional benefit of a larger system is related to a possible future study of the braiding phase of quasiholes, which appears more clearly when the quasiholes are sufficiently far apart. For this purpose, we project the operators into a space spanned by a group of low-energy single-particle states forming a narrow band for which the kinetic term is frozen out just analogously as in the LLL approximation in continuum. In the continuum limit $p/q \ll 1$, this lowest band is composed of $N_{\phi}$ lowest-energy states. LLL projected site operators are formed via defining $\tilde{c}_i^{\dagger} \equiv \sum_{k} \psi^k_i(\vec{\theta}) d_k^{\dagger}$, where $\vec{\theta}$ denotes the extra twist angles modifying magneto-periodic boundary conditions \cite{HafeziTorus} and $d_k^{\dagger}$ creates a particle in the $k$th lattice LLL state, $\psi^k_i(\vec{\theta})$ being its $i$th component. The projected interaction Hamiltonian then becomes

\begin{multline}
H_{I} =  \frac{U}{2}\sum_{i}\tilde{c}_i^{\dagger}\tilde{c}_i^{\dagger}\tilde{c}_i\tilde{c}_i = \frac{U}{2}\sum_{i}\sum_{k,l,m,n}d_{k}^{\dagger}d_{l}^{\dagger}d_{m}d_{n} \\ \times \psi^{k*}_{i}(\vec{\theta})\psi^{l*}_{i}(\vec{\theta})\psi^{m}_{i}(\vec{\theta})\psi^{n}_{i}(\vec{\theta}).
\label{eq:H_Interaction} 
\end{multline}

We also add to the model Hamiltonian (\ref{eq:Hofstadter Hamiltonian}) a repulsive impurity potential $V_{\rm imp} = Vn_i$, creating an offset with large enough strength $V>0$, to pin a quasihole on the $i$th site. Note that the impurity potential must also be projected into the LLL. Throughout our numerical calculations for an $L_x \times L_y$ lattice, we choose the flux density to be $\phi = 1/L_y$ to fit a full magnetic unit cell along the $y$ direction (for the chosen Landau gauge vector potential in the $x$ direction), which leads to $N_{\phi} = L_x$.

\begin{figure}[htbp]
	\includegraphics[width=\columnwidth]{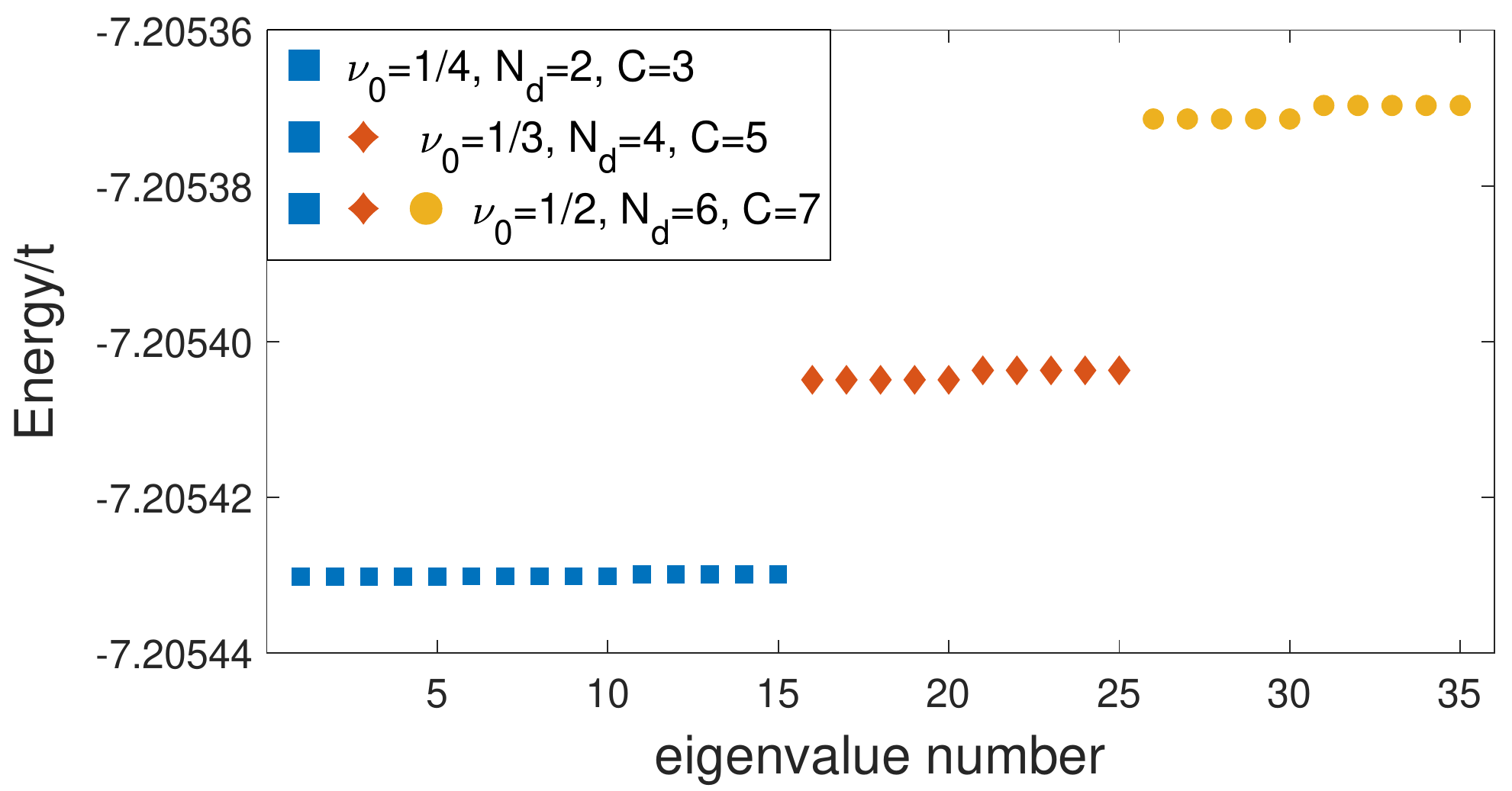}
	\caption{Energy spectrum for the quasihole states in the presence of hardcore interactions for $N = 2$ particles with $L_x = 10$, $L_y = 15$, $\phi = 1/15$. The gap $\sim 0.15t$ between the quasihole manifold and spurious states is much larger than the energy spread displayed here. In the legend, possible groupings of quasihole states based on the reference filling fraction $\nu_0$ (without the boson-fermion distinction corresponding to even-odd denominators of $\nu_0 = 1/r$; see text) are indicated by different combinations of symbols. $C$ is the Chern number of a subgroup. The grouping of the first five states for the possible $\nu_0 = 1/5$ reference filling is not shown as the splitting is very small compared to other groups and the Chern number cannot be calculated reliably. 
		\label{fig:spec}}
\end{figure}

\section{Grouping the quasihole states} 

For a bosonic system on a lattice with a filling fraction different from the Laughlin filling $\nu = 1/2$, there exists a manifold of nearly-degenerate quasihole states harboring delocalized quasiholes in the absence of a pinning potential \cite{realSpace,QHLiu}. These low-lying quasihole excitations are separated by a relatively large energy gap from spurious states. In addition to this gap, we numerically find smaller splittings within the manifold itself, which occur prominently at an integer multiple of $r$ when the filling fraction is $\nu = N/N_\phi = k/r$, $k$ and $r$ being coprime. Our degeneracy formula (\ref{DEG}) for continuum suggests that some of these splittings might mark the grouping of quasihole states with respect to a reference filling fraction as shown in Fig. \ref{fig:spec}. For instance, a system with $N=2$, $N_\phi = 10$ with the filling fraction $ \nu = 1/5 $ can be considered to host bosonic Laughlin quasihole states at different filling fractions such as $\nu_0=1/4$ with $N_d = 2$ ($\mathcal{D} = 15$) and $\nu_0=1/2$ with $N_d = 6$ ($\mathcal{D} = 35$) quasiholes. Since for a given number of particles and interaction strength the energy differences become inappreciable as the size of the system grows (see Appendix B to see an example of this effect), these energy differences may not be considered unambiguous observables and it is important to develop a method to distinguish between different types of quasihole excitations. Investigating, for instance, the particle entanglement spectrum might help one to make such a distinction \cite{RegnaultFCI,FCIexample,BernevigRegnault2012}. 

Here, we propose to label certain groups of states by their many-body Chern number \cite{ChernK,ChernManyB} (see Appendix C) in order to resolve the aforementioned ambiguity. In previous studies (e.g. \cite{HafeziTorus}) many-body Chern number has been shown to probe topological order by revealing whether a given collection of states reflects the properties of a Laughlin state of a certain filling fraction. In Table \ref{tab:table1}, we show the numerically calculated many-body Chern numbers $C$ of certain groups of low-lying states for several systems. Each group for a given $N$ and $N_\phi$ can be represented by a filling fraction of the Laughlin form $\nu_0 = 1/r$, $r$ being an integer (without the even-odd distinction) starting from $2$, plus a number $N_d = N_\phi-N/\nu_0$ of delocalized quasiholes so as to yield the physical filling $\nu = N/N_\phi$. As an odd-denominator filling fraction ($\nu_0 = 1/3$) appears in the Table as well, here is a good point to stress the fact that we do not claim that all these groups of states correspond to quasihole excitations of bosonic Laughlin states, for which only even denominators are allowed. Our motivation for considering all possible $\nu_0 = 1/r$ fractions is to show that odd-denominator fractions also fit in the combinatorial picture we present in the following (as in Figs.~\ref{fig:spec} and \ref{fig:COMB}), without too much speculating on the physical nature of the states. By localized or delocalized quasihole states, again we do not mean the usual bosonic Laughlin quasihole states, but just excitations over a state with a certain (even or odd) number of flux quanta per particle. 

\begin{table}[htbp]
	\caption{\label{tab:table1}
		Numerical many-body Chern numbers $C$ calculated using $\mathcal{D}$ lowest-energy states for systems of $N = 2,3$ hardcore bosons. $N_\phi$ is the number of flux quanta, $\nu_0$ is the reference filling fraction and $N_d$ is the corresponding number of delocalized quasiholes. Flux quantum per unit cell is $\phi = 1/12,1/10$ for $N = 2,3$, respectively. Degeneracy $\mathcal{D}$ is calculated using Eq.~(\ref{DEG}).}
	\begin{ruledtabular}
		\begin{tabular}{ccccccc}
			$N$&$N_\phi$&$\nu_0$&$N_d$&$\mathcal{D}$&$C,\mathcal{C}$&$C/\mathcal{D},\nu$\\ 
			\hline
			$2$&$5$&$1/2$&$1$&$5$&$2$&$2/5$\\
			\hline
			$2$&$6$&$1/3$&$0$&$3$&$1$&$1/3$\\
			$2$&$6$&$1/2$&$2$&$9$&$3$&$1/3$\\
			\hline
			$2$&$7$&$1/3$&$1$&$7$&$2$&$2/7$\\
			$2$&$7$&$1/2$&$3$&$14$&$4$&$2/7$\\
			\hline
			$2$&$8$&$1/4$&$0$&$4$&$1$&$1/4$\\
			$2$&$8$&$1/3$&$2$&$12$&$3$&$1/4$\\
			$2$&$8$&$1/2$&$4$&$20$&$5$&$1/4$\\
			\hline
			$3$&$6$&$1/2$&$0$&$2$&$1$&$1/2$\\
			\hline
			$3$&$7$&$1/2$&$1$&$7$&$3$&$3/7$\\
			\hline
			$3$&$8$&$1/2$&$2$&$16$&$6$&$3/8$\\
		\end{tabular}
	\end{ruledtabular}
\end{table}

Using the parameters $N$, $N_{\phi}$, and $N_d$, we find the corresponding degeneracy $\mathcal{D}$ through Eq.~(\ref{DEG}) and take $\mathcal{D}$ nearly-degenerate lowest-energy states of the system to calculate $C$ for this group of states. The most striking features observable in Table \ref{tab:table1} are: (i) Chern number $C$ equals the combinatorial factor $\mathcal{C}$ appearing in Eq.~(\ref{DEG}), (ii) Chern number per state $C/\mathcal{D}$ equals the physical filling fraction $\nu$ just as when $\nu$ itself is a Laughlin filling fraction. This result also suggests that all the states in the lowest-energy manifold share a common topological feature, namely $C/\mathcal{D}$. The appropriateness of this quantity in classifying states is apparent for the two systems displayed in the table with $\nu = 2/5$ and $\nu = 3/7$, which can both be interpreted as containing a single quasihole ($N_d = 1$) with respect to $\nu_0 = 1/2$, but each has a different $C/\mathcal{D}$ for the degenerate manifold. Next, we examine the role of pinning a quasihole in counting the degeneracy and suggest a connection between $C/\mathcal{D}$ and the charge of the pinned quasihole.

\section{The role of pinning} 

Adding an offset potential on a single site in the HH Hamiltonian (\ref{eq:Hofstadter Hamiltonian}) has been shown to be effective in pinning a single quasihole associated with an added flux quantum, visible as a density depletion around the site \cite{realSpace,QHLiu}. In general, the energy spectrum turns out to have a smoother variation in the presence of a pinning potential washing out the subgroup structure. We calculated the average spatial density $\expectationvalue{n_i}$ over the states that make up the nearly-degenerate manifold and computed the density depletion around the quasihole center via defining the missing density (or charge) with respect to its value in the absence of a pinning potential ($V = 0$) with one less flux quanta, which also corresponds to the almost uniform value of the numerical density calculated when a pinning potential is present, forming an analogy with the continuum case \cite{QHLiu,OnurTTN}:
\bea 
 Q_{\rho} \equiv \sum_{i}[\expectationvalue{n_i}_{V \neq  0} -\expectationvalue{n_i}_{V = 0}],
\label{eq:DEP}
\eea
where the sum is carried out for sites that are equidistant from the pinning site in a given radius $\rho$.

\begin{figure}[htbp]
	\includegraphics[width=\columnwidth]{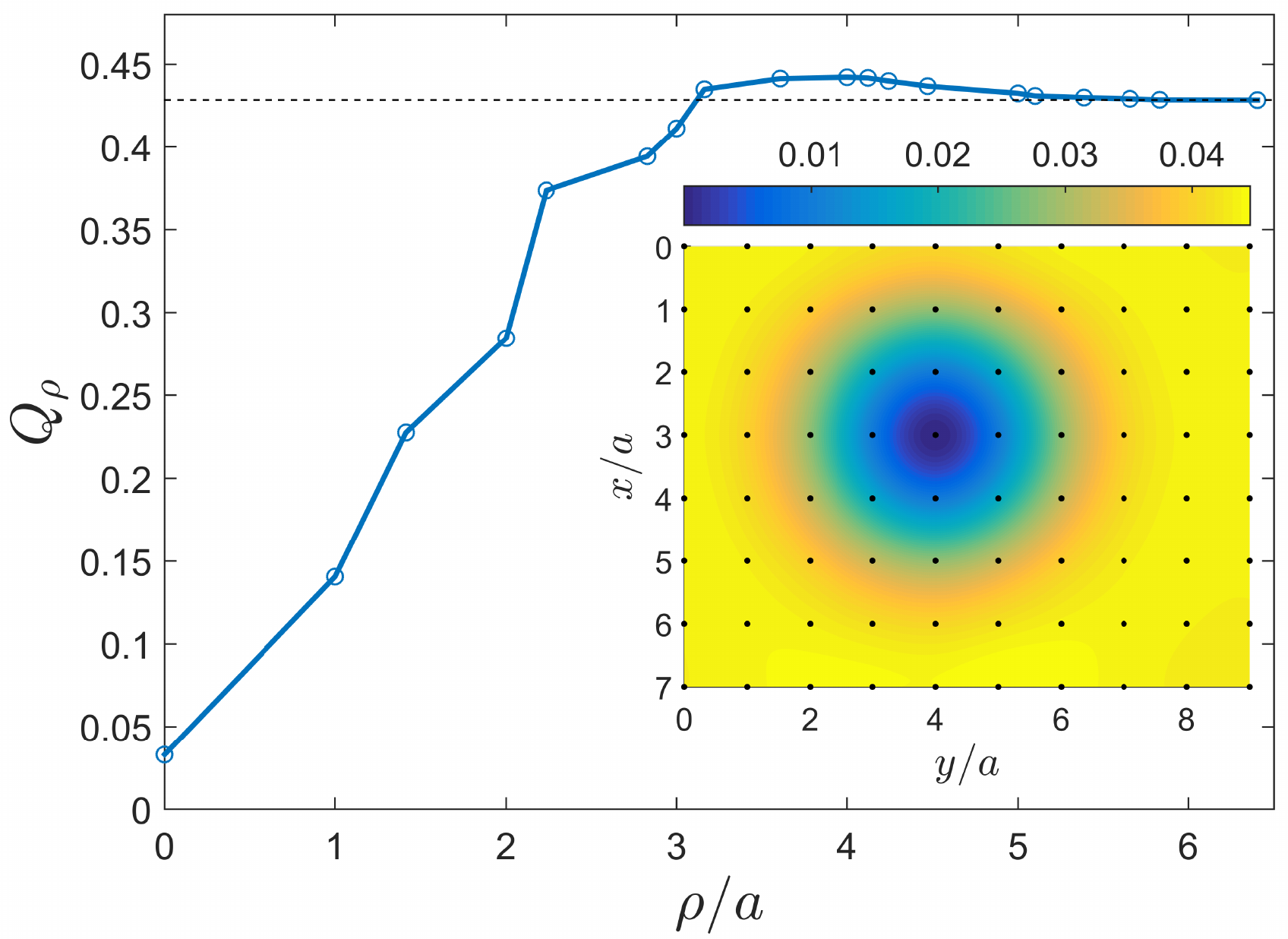}
	\caption{Density (charge) depletion profile $Q_{\rho}$ for a system of $N=3$ hardcore bosons with $V=4t$, $L_x = 8$, $L_y = 10$ and $\phi = 1/10$ (yielding $N_\phi=8$), as a function of the radial distance $\rho$ from the pinning site. Dashed line indicates the saturation level $N/(N_\phi-1)=3/7$ of the depletion occuring around $\rho \sim 3a$, which determines the quasihole size. The inset shows the site densities, interpolated for better visualization.
	\label{fig:depletion}}
\end{figure}

The fractional nature of the quasihole charge can be probed via the depletion profile. As seen in Fig.~\ref{fig:depletion}, the depletion saturates at a radius compatible with the quasihole size which can be extracted from the two-dimensional density. The saturation level then reveals the fractional charge. For the cases we numerically checked, this fractional charge always turned out to be equal to the filling $\nu=N/(N_\phi - N_{\rm loc})$ of a reference system without a pinning potential, containing a total of $N_\phi - N_{\rm loc}$ flux quanta, where $N_{\rm loc}$ is the number of localized flux quanta (or pinned quasiholes) in the actual system. We also observed that the number of states $D$ in the whole nearly-degenerate manifold of the actual system is the same as that of the reference system and can be given with Eq.~(\ref{DEG}) by considering only the unpinned flux quanta in the formula, through the replacement $N_\phi \rightarrow N_\phi - N_{\rm loc}$, as exemplified in Table \ref{tab:table2}. Moreover, for the systems given in the first two rows of the table, we calculated the many-body Chern number for the $7$-fold degenerate lowest band to be $3$, leading to the remarkable result that the Chern number per state $3/7$ in both cases is equal to the fractional charge of the pinned quasihole (Fig.~\ref{fig:depletion}), which is different from a simple Laughlin fraction $1/r$. Whether the quasihole (or more properly the localized density depletion) with charge $\nu = C/\mathcal{D}$ has any anyonic character with regard to braiding statistics (at least for simple bosonic fractions) is an interesting open problem to be pursued in a future work. In order to extract information as to the braiding phase of two quasiholes, one can either perform a real braiding operation with time-dependent potentials \cite{Kapit Non-Abelian, QHLiu, Nielsen2021} or take an indirect route which involves the comparison of the density profiles of one- and two-quasihole states \cite{OnurTTN}.

\begin{table}[htbp]
	\caption{\label{tab:table2}
		Numerical ground-state degeneracy $D$ for various systems obtained via exact diagonalization ($N=3$, hardcore) and LLL approximation ($N =4,5$). When there are $N_{\rm loc}$ localized quasiholes, each is centred at a lattice site through a pinning potential with the same strength. The number of delocalized quasiholes $N_d = (N_\phi - N_{\rm loc})-N/\nu_0$ is defined with respect to $\nu_0 = 1/2$. Symbols next to $D$ values are the same for systems with the same $N$ and $N_\phi - N_{\rm loc}$.}
	\begin{ruledtabular}
		\begin{tabular}{cccccc}
			$N$&$N_\phi$&$N_{\rm loc}$&$N_d$&$D$&$ $\\ 
			\hline
			$ 3 $ & $ 7 $ & $ 0 $ & $ 1 $ & $ 7 $ & \hspace{-1.5cm}$\ast$ \\
			\hline
			$ 3 $ & $ 8 $ & $ 1 $ & $ 1 $ & $ 7 $ & \hspace{-1.5cm}$\ast$ \\  
			\hline
			$ 4 $ & $ 9 $ & $ 0 $ & $ 1 $ & $ 9 $ & \hspace{-1.5cm}$\bullet$ \\ 
			\hline  
            $ 4 $ & $ 10 $ & $ 0 $ & $ 2 $ & $ 25 $ &\hspace{-1.5cm}$\triangle$\\			
			$ 4 $ & $ 10 $ & $ 1 $ & $ 1 $ & $ 9 $ & \hspace{-1.5cm}$\bullet$ \\
			\hline
			$ 4 $ & $ 11 $ & $ 0 $ & $ 3 $ & $ 55 $ & $ $\\ 
			$ 4 $ & $ 11 $ & $ 1 $ & $ 2 $ & $ 25 $ & \hspace{-1.5cm}$\triangle$\\  
			$ 4 $ & $ 11 $ & $ 2 $ & $ 1 $ & $ 9 $ & \hspace{-1.5cm}$\bullet$ \\
			\hline
			$ 5 $ & $ 11 $ & $ 0 $& $ 1 $ & $ 11 $ & \hspace{-1.5cm}$\square$\\
			\hline
			$ 5 $ & $ 12 $ & $ 0 $& $ 2 $ & $ 36 $ & $ $ \\ 
			$ 5 $ & $ 12 $ & $ 1 $& $ 1 $ & $ 11 $ &\hspace{-1.5cm}$\square$\\	  
		\end{tabular}
	\end{ruledtabular}
\end{table}

\section{Exact degeneracy in the Kapit-Mueller model}
Here, we briefly discuss whether the splittings we found in our numerical calculations can be a finite-size artifact, by considering the Kapit-Mueller model \cite{KapitMueller} for which the lowest single-particle band is exactly flat. Instead of the nearest-neighbor hopping of Eq.~(\ref{eq:Hofstadter Hamiltonian}), this model allows for an infinite-range hopping with properly decaying amplitudes \cite{Oktel2013}, which leads to the remarkable result that the lowest single-particle band of an infinite lattice is spanned by the continuum LLL functions sampled at lattice points. We calculated the energy spectra and many-body Chern numbers for several small Kapit-Mueller systems with high $\phi$, in the hardcore limit and using periodic boundary conditions. For all the cases we present in Table~\ref{tab:table3}, the many-particle groundstate manifold turned out to be exactly degenerate up to numerical precision even in the presence of a pinning potential. Moreover, the degeneracy and the many-body Chern number for this manifold were found to be as expected, conforming to the formulas presented in previous sections.

\begin{table}[htbp]
	\caption{\label{tab:table3}
		Numerical ground-state degeneracy $D$ and many-body Chern number $C$ for Kapit-Mueller systems of $N = 2,3$ hardcore bosons with respective flux quanta per unit cell $\phi = 2/5,1/3$. The lattice size $L_x = L_y = 5,6$ determines the total number of flux quanta as $N_\phi = 10,12$. The number $N_{\rm loc} = 1 (0)$ signifies that a quasihole is (not) localized with a pinning potential. The number of delocalized quasiholes $N_d = (N_\phi - N_{\rm loc})-N/\nu_0$ is defined with respect to $\nu_0 = 1/2$.}
	\begin{ruledtabular}
		\begin{tabular}{ccccccc}
			$N$&$N_\phi$&$N_{\rm loc}$&$N_d$&$D$&$C$&$C/D$\\ 
			\hline
			$2$&$10$&$0$&$6$&$35$&$7$&$1/5$\\
			$2$&$10$&$1$&$5$&$27$&$6$&$2/9$\\
			\hline
			$3$&$12$&$0$&$6$&$112$&$28$&$1/4$\\
			$3$&$12$&$1$&$5$&$77$&$21$&$3/11$\\			
		\end{tabular}
	\end{ruledtabular}
\end{table}

The exact degeneracy appearing in these systems, which are even smaller than the ones investigated in previous sections with the same boundary conditions and therefore could in principle be prone to the same finite-size issues, suggests that the small splittings we obtained for the Hofstadter-Hubbard model in the continuum limit might be ascribed to the nearest-neighbor tight-binding nature of the model rather than to the small size of the systems we investigated. From this standpoint, we may argue that the nearest-neighbor hopping model might help us better understand the competing many-body phases in the continuum by lifting the exact degeneracy and leading to an intricate grouping structure of states.

\section{Possible generalization of the combinatorial scheme} 

Until this point, we have treated energy splittings that correspond only to certain groups of quasihole excitations (as in Fig.~\ref{fig:spec}) and provided a simple combinatorial factor ($\mathcal{C}$) for the Chern number of subgroups. This factor can be interpreted as the number of ways to distribute $N_d$ indistinguishable objects into $N$ distinguishable boxes. Based on numerical calculations for larger systems (without pinning), we observed that there exist extra splittings and correspondingly new groupings in the degenerate quasihole manifold, which can be accounted for by a more general combinatorial scheme. 

\begin{figure}[htbp]
	\includegraphics[width=\columnwidth]{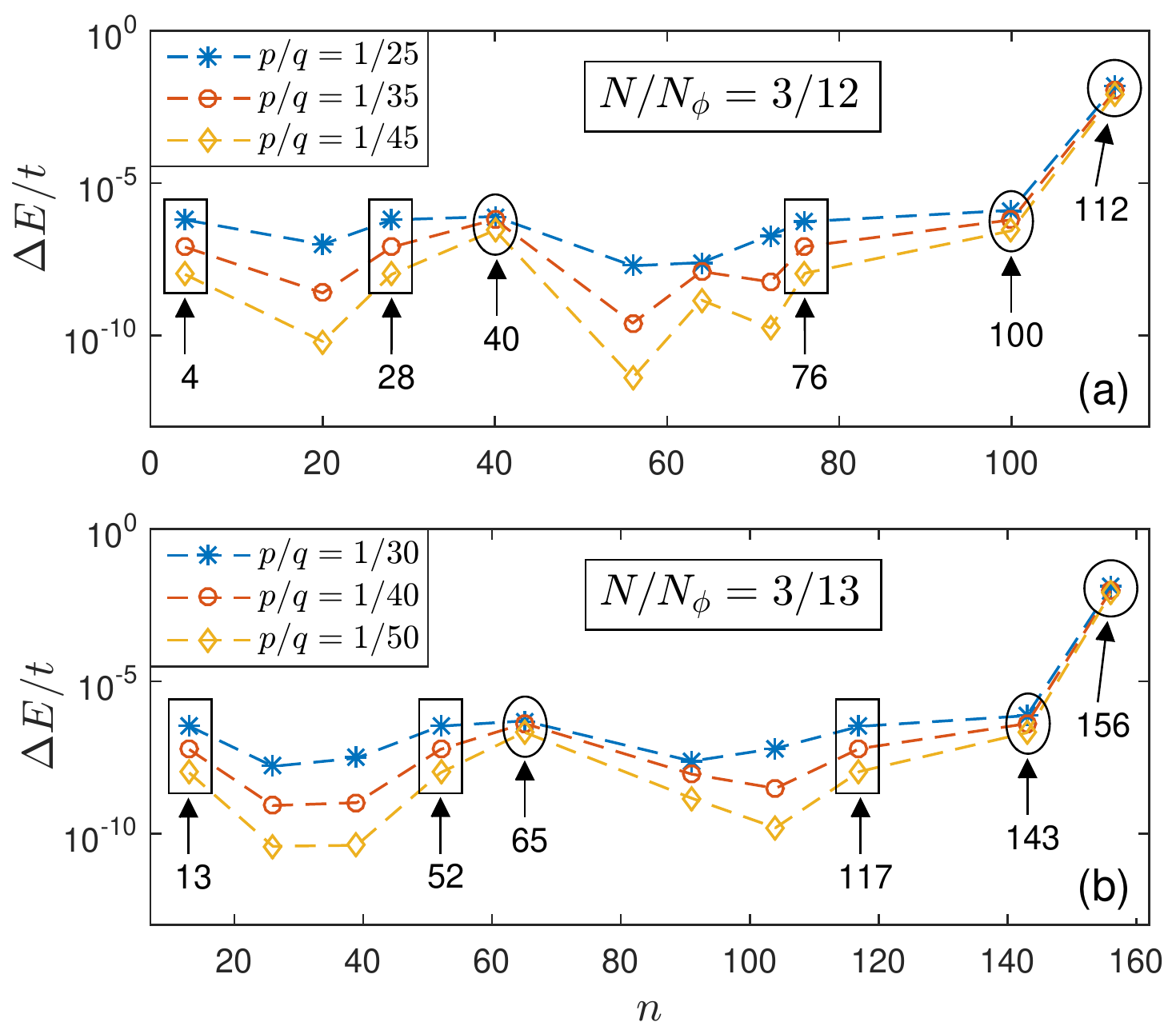}
	\caption{For systems with $\nu = N/N_\phi = 3/12$ and $\nu = N/N_\phi = 3/13$ at various $\phi = p/q \ll 1$ and fixed $U/t$, the largest ten of the energy differences $\Delta E(n) = E_{n+1}-E_n$ between consecutive states of increasing energy are displayed in logarithmic scale. 
		\label{fig:COMB}}
\end{figure}

In Fig.~\ref{fig:COMB}, we display the largest ten mini-gaps between consecutive states of the lowest-energy manifold for two systems in the continuum limit with various $\phi = p/q \ll 1$. The largest gap (delineated by a circle) separating the whole nearly-degenerate manifold from spurious states is quite robust. It is also clearly seen that certain mini-gaps (delineated by ovals and rectangles) close in a similar fashion as $\phi$ gets smaller. Upon scrutiny, we realized that all these relatively slowly closing gaps mark groups of states with degeneracy $\tilde{D} = \tilde{C}/\nu$, where $\nu = N/N_{\phi}$ and $\tilde{C}$ is a new combinatorial factor encompassing the previously found $\mathcal{C}$. Now, $\tilde{C} \equiv (N,N_d,m)$ corresponds to the number of ways to distribute $N_d$ indistinguishable objects into $N$ distinguishable boxes each of which can accommodate a maximum of $m$ objects. For $m = N_d$, $\tilde{C}$ equals $\mathcal{C}$. In order to exemplify this new counting, we focus on the system of Fig.~\ref{fig:COMB}(a) with $N=3$ and $N_\phi=12$. This system can be thought to possess excitations at different filling fractions: $\nu_0 = 1/2$ with $N_d = 6$, $\nu_0 = 1/3$ with $N_d = 3$, and $\nu_0 = 1/4$ with $N_d = 0$. The degeneracy of the groups is then obtained by computing $\tilde{C}$ for certain $m$ as follows 

\begin{align*}
\hline
\hline
    &\tilde{C}    &\tilde{D} = \tilde{C}&/\nu \\
        \hline
	(N=3,N_d =6&,m=6) =28  &112 \\
	(N=3,N_d =6&,m=5) =25  &100 \\
	(N=3,N_d =6&,m=4) =14  &76 \\
	(N=3,N_d =3&,m=3) =10  &40 \\
	(N=3,N_d =3&,m=2) =7   &28 \\
	(N=3,N_d =0&,m=0) =1   &4 \\
	\hline
	\hline	
\end{align*} 

Results for the system of Fig.~\ref{fig:COMB}(b) follow analogously. We have also encountered extra slow-closing gaps in numerical results for various $N = 4$ systems, for which we could not find a distinct pattern. Although we were not able to perform a reliable Chern number calculation using the LLL projection method, the relation $\nu = \tilde{C}/\tilde{D}$ suggests that this new combinatorial factor $\tilde{C}$ can be thought to be the many-body Chern number of the $\tilde{D}$-fold nearly-degenerate states, by analogy with our previous findings.

\section{Conclusion} 

We presented a compact expression for the degeneracy of Laughlin quasihole states at different filling fractions on a torus, which can be written as the inverse physical filling fraction times a simple combinatorial factor. Based on numerical many-body Chern number calculations for a lattice model, we observed that the combinatorial part equals the Chern number of certain subgroups of states suggesting that all states in the lowest-energy manifold can be characterized by the same Chern number per state, which also equals the physical filling fraction. In systems containing pinned quasiholes, we showed that the quasihole charge is controlled by the number of unpinned flux quanta and is equal to the Chern number per state as well. 

In order to study larger systems with slightly larger number of particles and flux quanta in the continuum limit, which is characterized by small flux quanta per unit cell, we employed a lowest-band approximation. As one approaches the continuum limit of an exactly flat manifold of states, the energy-splittings get smaller. By examining the closing structure of these mini-gaps, we also proposed a more general combinatorial scheme to account for additional groupings that do not correspond to quasihole states at simple $\nu_0 = 1/r$ fractions. We briefly touched upon the possibility of these energy splittings being a finite-size artifact as well, by studying the Kapit-Mueller model of long-range hoppings for small systems. Having found the many-particle ground state manifold to be exactly flat for these small systems, we speculated that the origin of the small splittings we encounter, with their specific grouping structure, might be the short-range hopping rather than the smallness of the system. The grouping structure of the nearly-degenerate states we identified and the compact form of our degeneracy formula with its relation to some basic topological invariants can be considered a step forward towards a better microscopic understanding of the system. As a next step, we will focus on various FCI systems and the braiding properties of quasiholes in the direction of expanding our present findings.

\begin{acknowledgments} 
The authors thank I. Carusotto for inspiring discussions. This work has been supported through the T\"UB\.{I}TAK-CNR International Bilateral Cooperation Program 2504 (project no.~119N192). R.~O.~U.  acknowledges support through the T\"UBA-GEB\.{I}P Award of the Turkish Academy of Sciences.
\end{acknowledgments}

\appendix

\section{Proof of the generalized counting formula}

The counting procedure for the total number of quasihole states of a $\nu = 1/r $ Laughlin state of $ N$ particles in $ N_\phi $ orbitals (number of flux quanta) on the torus can be described in two steps \cite{RegnaultFCI}:

\begin{itemize}
	\item Enumerate the number of ways of occupying $ N_\phi $ orbitals with $ N $ particles with respect to the constraint that the configurations obey the $ (1,r) $ generalized Pauli principle on the sphere \cite{JACK}. 
	\item Discard occupation configurations which violate the torus boundary conditions.
\end{itemize}

\noindent Thus the problem of degeneracy counting can actually be analyzed as two separate counting problems.

\subsection{Quasihole Counting on the Sphere}

Given $ N_\phi $ orbitals and $ N $ particles we must enumerate all occupation configurations such that no more than one particle exists in $ r $ consecutive orbitals, according to the $ (1,r) $ generalized Pauli principle.

Let $x_0,x_1,\dots,x_N$ represent the number of unoccupied orbitals (zeros) between the consecutive occupied orbitals (ones). For instance, $x_0$ is the number of zeros to the left side of the first occupied orbitals and $x_1$ is the number of zeros between the first and second occupied orbitals, and so forth: 

\[\underbrace{0}_{x_0}1\underbrace{000}_{x_1}1\underbrace{000}_{x_2}1\underbrace{0000}_{x_3}1\dots1\underbrace{000}_{x_N}\] \\
We know there are a total of $N_\phi-N$ zeros indicating unoccupied orbitals. Thus we have 

\begin{equation} \label{eq:int}
	x_0+x_1+\dots+x_N=N_\phi-N.
\end{equation}

\noindent Furthermore, the constraint that no more than one particle exist in $ r $ consecutive orbitals implies that for $i\in\{1,2,\dots,N-1\}$, configurations must conform to the condition $x_i\geq r-1$. Therefore, the problem reduces to the number of integer solutions to (\ref{eq:int}) with the constraint $x_i\geq r-1$ for $i\in\{1,2,\dots,N-1\}$: 

%\flushleft
\begin{eqnarray}
\begin{drcases}x_0+x_1+\dots+x_N=N_\phi-N\\ x_0\geq0\\ x_1\geq r-1\\ x_2\geq r-1\\ \vdots\\ x_{N-1}\geq r-1 \\ x_N\geq0 \end{drcases}\label{eq:Set_1}
\end{eqnarray}

%\flushleft
In order to have the constraint included inside the integer equation, we make a change of variables in (\ref{eq:Set_1}):  $a_0=x_0, a_N=x_N, a_i=x_i+1 -r$. Then the integer equation takes the following form

%\flushleft
\begin{eqnarray}
\begin{drcases}a_0+a_1+\dots+a_N=N_\phi-Nr+r-1\\ a_0\geq0\\ a_1\geq0\\ a_2\geq0\\ \vdots\\ a_{N-1}\geq0 \\ a_N\geq0 \end{drcases}
\end{eqnarray} 

\noindent
with the solution,
\begin{equation*}
	\boxed{\mbox{Number of integer solutions} =\binom{N+n}{n}}
\end{equation*}
where $ n \equiv N_\phi-Nr+r-1 $.

\subsection{Quasihole Counting on the Torus}

We now calculate the number of configurations that violate the torus boundary conditions (TBCs) for a given $ r $. TBCs are realized through applying the 1D-chain boundary condition to the $ N_\phi $-orbital space, i.e., the $ i $th orbital is identified with the $( i+N_\phi )$th one.

\vspace{2mm}
\begin{tcolorbox}
	\textbf{Example:} Boundary condition violating patterns for quasihole excitations of $ \nu = \frac{1}{3} $
	\vfill
	\[
	1\tikzmark{x1}00\underline{\hbox to 6cm{\hss (1,3) admissible-patterns\hss}}001\myline[dashed]\tikzmark{x2}1 	\tikz[overlay,remember picture]
	{\draw[->,square arrow] (x1.south) to (x2.south);}
	\]
	\vspace{1mm} 
	
	\[
	1\tikzmark{x1}00\underline{\hbox to 6cm{\hss (1,3) admissible-patterns\hss}}0010\myline[dashed]1\tikzmark{x2}0 	\tikz[overlay,remember picture]
	{\draw[->,square arrow] (x1.south) to (x2.south);}
	\]
	\vspace{1mm} 
	
	\[
	0\tikzmark{x1}100\underline{\hbox to 6cm{\hss (1,3) admissible-patterns\hss}}001\myline[dashed]0\tikzmark{x2}1 	\tikz[overlay,remember picture]
	{\draw[->,square arrow] (x1.south) to (x2.south);}
	\]
	\vspace{1mm}
	
	Above configurations are all valid on the sphere but not so when TBCs are applied, thus any inner (1,3) admissible patterns between the edge patterns such as above cannot be realized on the torus and must be discarded.  
	
\end{tcolorbox}
\vspace{5mm}

As it can be seen from the example, TBCs lead to a reduction in the degeneracy; therefore, in order to obtain the correct number we must count invalid configurations and discard them. We will show the counting for the $ \nu=1/3 $ case and generalize it for a given $ \nu=1/r $.
\vspace{5mm}

\begin{align*} 	
	\textbf{(a)}\:\:\: &100\text{\underline{\hbox to 6cm{\hss (1,3) admissible-patterns\hss}}}001 \\
&\text{Unoccupied }= N_\phi - N -4 \\
&\text{Occupied }= N-2	
\end{align*}

Repeating similar calculations, as we carried out for the sphere, to count the inner (1,3) admissible-configurations with fixed edge patterns, we obtain

%\flushleft
\begin{eqnarray}
\begin{rcases}x_0+x_1+\dots+x_{N-2}=N_\phi-N-4\\ x_0\geq0\\ x_1\geq 2\\ x_2\geq 2\\ \vdots\\ x_{N-3}\geq 2 \\ x_{N-2}\geq0 \end{rcases}\label{eq:Set_2}
\end{eqnarray}

%\flushleft
As we did in the sphere counting, we make a change of variables in (\ref{eq:Set_2}) to include the constraint inside the integer equation:  $a_0=x_0, a_N=x_N, a_i=x_i-2$, which yields

%\flushleft
\begin{eqnarray}
\begin{rcases}a_0+a_1+\dots+a_{N-2}=N_\phi-3N+2\\ a_0\geq0\\ a_1\geq0\\ a_2\geq0\\ \vdots\\ a_{N-3}\geq0 \\ a_{N-2}\geq0 \end{rcases}
\end{eqnarray} 

%\flushleft
with the solution,
\begin{equation*}\label{eq:sp1}
	\boxed{\mbox{Number of integer solutions} =\binom{N+n-2}{n}}
\end{equation*}

\vspace{5mm}
\begin{align*} 	
	\textbf{(b)}\:\:\: &100\text{\underline{\hbox to 6cm{\hss (1,3) admissible-patterns\hss}}}0010 \\ &\text{Unoccupied }= N_\phi - N -5 \\
&\text{Occupied }= N-2	
\end{align*}

In this case, calculations are same as in the above case (a) except that the number of unoccupied orbitals changes and the result is
\begin{equation*}\label{eq:sp2}
	\boxed{\mbox{Number of integer solutions} =\binom{N+n-3}{n-1}}
\end{equation*}
\vspace{5mm}

Due to inversion symmetry, 0100\text{\underline{\hbox to 6cm{\hss (1,3) admissible-patterns\hss}}}001 generates the same number of configurations as (b). Then the total number of degenerate states for the quasihole excitations of $ \nu=1/3 $ with TBCs is

\[ \mathcal{D_{1/3}} = \binom{N+n}{n} - \binom{N+n-2}{n}-2\binom{N+n-3}{n-1}. \]

For a generic fraction $ \nu=1/r $, we see by inspection that there exist $ r-1 $ independent boundary-condition violating terms which need to be multiplied by appropriate constant factors due to inversion symmetry in order to find the total number of disallowed configurations. The proposed final formula for the degeneracy is
\begin{widetext}
\begin{multline}
	\mathcal{D}_{1/r} = \binom{N+n}{n} - \binom{N+n-2}{n}-2\binom{N+n-3}{n-1} - \dots - (r-1)\binom{N+n-r}{n-r+2} \\
	=(rN+n-r+1)\frac{(N+n-r)!}{N!(n-r+1)!}, \label{eq:Main_eq}
\end{multline}
\end{widetext}
where $ n \equiv N_\phi-Nr+r-1 $. We will prove the last equality by induction. We obtained the right-hand side combinatorial form of the degeneracy by guessing from the numerical degeneracies we encountered in both our calculations and some other works, e.g. \cite{RegnaultFCI,FCIexample,BernevigRegnault2012}. Later we have become aware of Ref. \cite{ThinTorusCount}, which contains essentially the same formula (in its footnote 49) derived in the thin-torus limit.
\vspace{5mm}

\textbf{Proof of Eq.~(\ref{eq:Main_eq}):} One can easily show that Eq.~(\ref{eq:Main_eq}) is true for $r = 2$. Assuming that the equation holds true for $ r = k $:
\begin{widetext}
\begin{equation}
\binom{N+n}{n} - \binom{N+n-2}{n} - \dots - (k-1)\binom{N+n-k}{n-k+2}=(kN+n-k+1)\frac{(N+n-k)!}{N!(n-k+1)!},\label{eq:r_eq_k}
\end{equation}
\end{widetext}
we will show that it holds true for $r = k+1$:
\begin{widetext}
\begin{multline}
 \binom{N+n}{n} - \binom{N+n-2}{n} - \dots - (k-1)\binom{N+n-k}{n-k+2} - k\binom{N+n-k-1}{n-k+1} \\ =(kN+N+n-k)\frac{(N+n-k-1)!}{N!(n-k)!}.\label{eq:r_eq_kplus1}
\end{multline}
\end{widetext}

Inserting Eq.~(\ref{eq:r_eq_k}) in the left-hand side of Eq.~(\ref{eq:r_eq_kplus1}) and rearranging terms yield

\begin{widetext}
\begin{align*}
	(kN+n-k+1)\frac{(N+n-k)!}{N!(n-k+1)!}-k\frac{(N+n-k-1)!}{(n-k+1)!(N-2)!} &= \\ 
	(kN+n-k+1)\frac{(N+n-k)(N+n-k-1)!}{N!(n-k+1)!} -k\frac{N(N-1)(N+n-k-1)!}{N!(n-k+1)!} &=\\
	\frac{(N+n-k-1)!}{N!(n-k)!}\Bigg[\frac{(kN+n-k+1)(N+n-k)-kN(N-1)}{n-k+1}\Bigg]&=\\
	%\frac{(N+n-k-1)!}{N!(n-k)!}\Bigg[\frac{N-k+n+Nn-2kn-Nk^2+k^2+n^2+Nkn}{n-k+1}\Bigg]&=\\
	\frac{(N+n-k-1)!}{N!(n-k)!}\Bigg\{\frac{(n-k+1)(N+n-k)+kN\big[N+n-k-(N-1)\big]}{n-k+1}\Bigg\}&=\\
	\frac{(N+n-k-1)!}{N!(n-k)!}\Bigg[\frac{\cancel{(n-k+1)}(N+n-k+kN)}{\cancel{(n-k+1)}}\Bigg]&=\\
	(kN+N+n-k)\frac{(N+n-k-1)!}{N!(n-k)!}&=\\
	&&\blacksquare
\end{align*}
\end{widetext}

Therefore, for a given $ \nu=1/r $ we can obtain the number of degenerate quasihole excitations, and we can have it in a form that only depends on system parameters by using the definitions made previously. Since $ n=N_\phi-Nr+r-1 $, 

\begin{align*}	
	\mathcal{D}_{1/r} &= (rN+n-r+1)\frac{(N+n-r)!}{N!(n-r+1)!} \\
	&=N_\phi\frac{(N+N_\phi-Nr-1)!}{N!(N_\phi - Nr)!} \\
	&\boxed{\mathcal{D}(N,N_d,N_\phi)=\frac{(N_d + N -1)!}{N_{d}!(N-1)!}\frac{N_\phi}{N}},
\end{align*}
where $ N_\phi = Nr + N_d $, $ N_d $ being the number of delocalized quasiholes in the system.

\section{Energy gaps and the system size}

\begin{figure}[htbp]
	\includegraphics[width=\columnwidth]{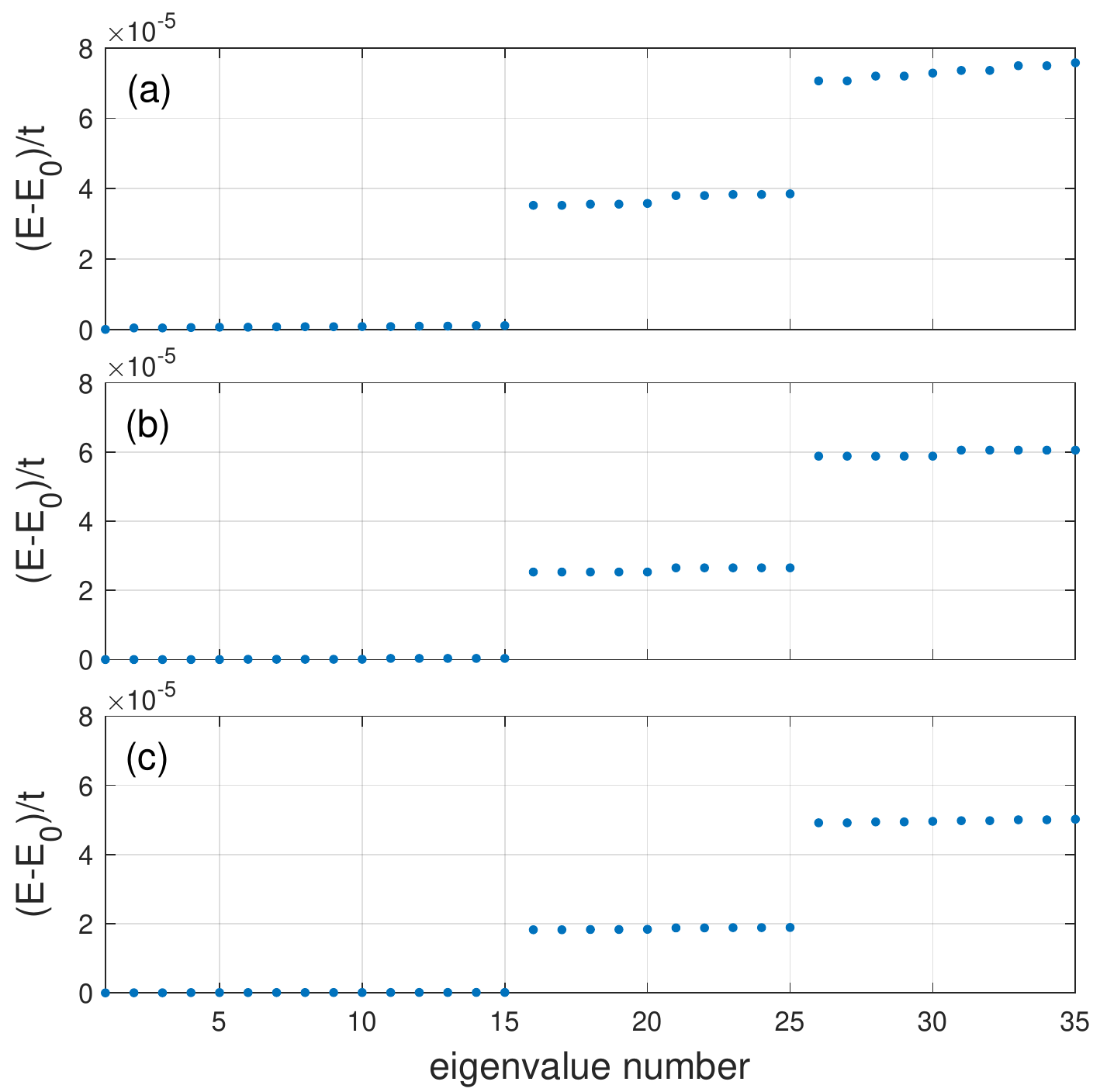}
	\caption{Diminishing of the energy gaps as the system size increases. Hardcore interactions are assumed for $N = 2$ particles with lattice size $L_x = 10$, and (a) $L_y = 14$, (b) $L_y = 15$, (c) $L_y = 16$. In each case, flux is taken to be $\phi = 1/q$ with $q = L_y$ and $E_0$ is the lowest energy.
		\label{fig:energygap}}
\end{figure}

In this appendix, we exemplify the dependence of the small energy gaps marking different groups of states on the system size. This effect is already implicitly contained in Fig.~\ref{fig:COMB} of the main text, where for fixed interaction strength and number of particles it is seen that the mini-gaps close as one approaches the continuum limit of smaller flux quanta per unit cell. As we mentioned  below Eq.~(\ref{eq:H_Interaction}) of the main text, in our numerical modeling, we choose to fit full magnetic unit cells along any direction of the lattice in order to avoid possible boundary effects. As a result, while going to the continuum limit by decreasing the flux quanta per unit cell $\phi = p/q$ through increasing $q$ ($= L_y)$, the lattice size increases automatically. This relation also ensures that $L_x$ flux quanta are contained in the whole lattice at all times. In Fig.~\ref{fig:energygap}, we show this effect for three nearby $q$ values (panel(b) has the same parameters as in Fig.~\ref{fig:spec}) for a system of two hardcore bosons. It is apparent that the energy gaps decrease with increasing $q = L_y$.

\section{Many-body Chern number calculation}

Here, for completeness, we briefly outline the method of Refs. \cite{ChernK,LLL,ChernManyB} for the many-body Chern number calculation. In this method, the Berry curvature is expressed in a discretized parameter space of twisted angles $\vec{\theta} = (\theta_x,\theta_y)$ of the torus boundary conditions \cite{Niu}. In our numerical calculations twisted angles are discretized as $ \theta_i = \frac{2\pi}{N_\theta}n_i $ ($ n_i = 1,\dots, N_\theta; i=x,y $), where $ N_\theta $ tunes the fineness of the mesh. The discretized Berry curvature is defined as $\mathcal{F(\vec{\theta})}\!=\!\text{log}[U_x(\vec{\theta})U_y(\vec{\theta}+\delta_x)U_x(\vec{\theta}+\delta_y)^{-1}U_y(\vec{\theta})^{-1} ]$, where the link variables $ U_i(\vec{\theta})$ are expressed as
\bea
U_i(\vec{\theta}) = \frac{\text{det}[\Phi^{\dagger}(\vec{\theta})\Phi(\vec{\theta} + \delta_i)]}{|{\text{det}[\Phi^{\dagger}(\vec{\theta})\Phi(\vec{\theta} + \delta_i)]|}};\:\: i = x,y.
\label{eq:U1}
\eea
Here, $ \Phi(\vec{\theta}) $ stands for the multiplet matrix $ \Phi(\vec{\theta})= (\ket{G_1(\vec{\theta})},\dots,\ket{G_\mathcal{D}(\vec{\theta})}) $ formed by $ \mathcal{D} $-fold degenerate states for which we want to calculate the Chern number and $\delta_x = (\frac{2\pi}{N_\theta},0)$, $\delta_y = (0,\frac{2\pi}{N_\theta})$ are the incremental units in parameter space. Calculation of the link variable $ U_i(\vec{\theta})$ involves finding the overlap between states at different twisted angles corresponding to different boundary conditions. From this construct, the Chern number can be calculated by taking the sum over all possible twisted angles in the mesh:

\bea
\mathcal{C} = \frac{1}{2\pi i} \sum_{\vec{\theta}} \mathcal{F(\vec{\theta})}.
\eea

\end{document}